\documentclass[a4paper,11pt]{article}
\usepackage{jheppub} 
\usepackage{lineno}

\arxivnumber{} 

\title{\boldmath The Effect of Higher Harmonics On Gravitational Wave Dark Sirens}

\author[a]{Jian-Dong Liu}
\author[b,a,c,d]{Wen-Biao Han}
\author[e,b]{Qianyun Yun} 
\author[b]{and Shu-Cheng Yang}
\affiliation[a]{Hangzhou Institute for Advanced Study, University of Chinese Academy of Sciences, \\Hangzhou 310124, China}
\affiliation[b]{Shanghai Astronomical Observatory, Chinese Academy of Sciences, \\Shanghai 200030, China}
\affiliation[c]{School of Astronomy and Space Science, University of Chinese Academy of Sciences, \\Beĳing 100049, China}
\affiliation[d]{Taĳi Laboratory for Gravitational Wave Universe (Beĳing/Hangzhou), University of Chinese Academy of Sciences, \\Beĳing 100049, China}
\affiliation[e]{School of Physics and Astronomy, Shanghai Jiao Tong University 800 Dongchuan RD.,Minhang District, \\Shanghai 200240, China}

\emailAdd{wbhan@shao.ac.cn}

\abstract{The gravitational wave (GW) signal from the merger of two black holes can serve as a standard sirens for cosmological inference. However, a degeneracy exists between the luminosity distance and the inclination angle between the binary system’s orbital angular momentum and the observer’s line of sight, limiting the precise measurement of the luminosity distance. In this study, we investigate how higher harmonics affect luminosity distance estimation for third-generation (3G) GW detectors in binary black hole mergers. Our findings demonstrate that considering higher harmonics significantly enhances distance inference results compared with using only the (2,2) mode. This improved accuracy in distance estimates also strengthens constraints on host galaxies, enabling more precise measurements of the Hubble constant. These results highlight the significant influence of higher harmonics on the range estimation accuracy of 3G ground-based GW detectors.}


\begin{document}
\maketitle
\flushbottom

\section{Introduction}\label{sec:1}

The detection of gravitational waves produced by the merger of binary black holes (BBH) marks a milestone in modern astrophysics and is one of its most groundbreaking achievements. In 2015, LIGO announced the first direct capture of a gravitational wave event, GW150914 \cite{2016PhRvX...6d1015A,Abbott_2016_150914,Abbott_2016BBH,2015CQGra..32g4001L}, which marked the beginning of gravitational wave astronomy. Two years later, during the second observation run (O2, Observation Run 2) \cite{2015CQGra..32b4001A}, the LIGO/Virgo joint detection of GW170817 \cite{Abbott_2017,Abbott_2019,Abbott_2017a}, a binary neutron star system with an electromagnetic counterpart, is historic and initiated a new era in multi-messenger astronomy. In the future, third-generation ground-based gravitational wave detectors such as Cosmic Explorer (CE) \cite{2019BAAS...51g..35R,Hall_2021CE} and Einstein Telescope (ET) \cite{Maggiore_2020ET,2012CQGra..29l4013S,2010CQGra..27s4002P} will be used, which will increase the sensitivity by at least two orders of magnitude compared with current detectors. The extension of the low-frequency cutoff to 1 Hz indicates a significant increase in the detection rate of gravitational wave signals. Simultaneously, we will be able to detect signals from higher redshift regions. For lower frequency gravitational wave sources, signals will be captured using space gravitational wave detectors such as LISA \cite{caldwell2019astro2020_lisa}, Taiji \cite{2020IJMPA..3550075R}, and TianQin \cite{2021PTEP.2021eA107M}. With the operation and networking of these telescopes, the field of gravitational wave astronomy research will be greatly expanded.

The discovery of GW170817 was a landmark. The LIGO/VIRGO used gravitational wave detectors to independently measure its luminosity distance, By combining this information with the redshift of its host galaxy, NGC 4993, this signal functioned as a bright sirens for the first time, helping to calculate the Hubble Constant by $H_{0} = 69^{+17}_{-8}$ km/s/Mpc \cite{2019NatAs...3..940H, Fishbach_2019}. Just like GW170817, when a gravitational wave event has an electromagnetic counterpart, we can identify the host galaxy through multi-messenger observations of its electromagnetic counterpart. By doing this, we can independently obtain the source’s redshift and use it to infer cosmological parameters, which is called a ``bright sirens" \cite{Abbott_2021}. However, so far, most detected signals have been from binary black hole mergers, which lack electromagnetic counterparts. This absence makes it challenging to identify the host galaxy and determine its redshift. In this scenario, the redshift information of the source can be obtained through statistical analysis of galaxy catalogues, known as ``dark sirens" \cite{Leandro_2022,Finke_2021,Chen_2018,nissanke2013determining}. In addition to measuring the Hubble Constant, standard sirens have a wealth of cosmological applications, such as limiting equations of state for dark energy \cite{Arabsalmani_2013}, distinguishing between dark energy, modifying gravitational theory \cite{Cutler_2009}, and studying modifying gravitational theory \cite{Belgacem_2019}.

Currently, there are many methods for determining the redshift of a source without an electromagnetic counterpart. Among them, using gravitational wave detectors to localise them and identify their host galaxy is a main approach for measuring the redshift of dark standard sirens \cite{2012JPhCS.363a2052N,MacLeod_2008}. By acquiring the right ascension, declination, and luminosity distance of the source and employing star catalogs to filter out host galaxies within the target region, we can indirectly obtain the redshift information of the source by weightening the redshifts of all these host galaxies. Then, combined with the flat $\Lambda$CDM model, the Hubble constant can be derived. It is worth mentioning that GW170814 was the first event where this method was employed to calculate the Hubble Constant. Using the localisation results of this event, along with the survey data of The Dark Energy Survey Collaboration (DES), the Hubble constant was limited to $H_{0} = 75^{+40}_{-32}$km/s/Mpc \cite{DES:2019ccw}.

The uncertainty in inferring cosmological parameters from ``dark sirens" depends on errors in the luminosity distance ($d_{L}$) and redshift ($z$), where $d_{L}$ is degenerate with the inclination angle of the orbit and $z$ error is constrained by the localisation of the host galaxy. The degeneracy between orbital inclination and luminosity distance arises because when the binary system is near face-on or face-off, the contributions of plus and cross polarizations to the overall gravitational wave amplitude are nearly equal, with their relative differences being less than 1\% \cite{Usman_2019},this leads us to struggle in distinguishing their contributions to the total amplitude. Therefore, we use a waveform template with higher harmonics. The amplitudes of the higher harmonics vary with different polarization angles \cite{Mills_2021}. This allows us to distinguish the contributions of plus and cross polarizations to the total amplitude, thus breaking the degeneracy.

The key to obtaining redshift information from galaxy catalogues lies in the restriction on the number of host galaxies. Currently, according to the latest results of the third LIGO-Virgo-KAGRA gravitational wave source analysis, when using galaxy catalogues in the k-band to search for host galaxies, the order of magnitude of the number of host galaxies for most sources is approximately $10^{4}$ , while only a few sources having fewer than 10 identified hosts. In addition, for most sources, the probability of host galaxy inclusion in the catalogues is not high \cite{LIGOScientific:2021aug}. To enhance the identification capability of host galaxies in the catalogue, we can consider introducing higher harmonics in future data analysis. For Massive black holes (MBHs), the introduction of higher harmonics can improve the identification capability by more than 70\% compared with only considering the (2,2) mode \cite{gong2023including}.

In this study, we employ two methods for investigation: one introduces higher harmonics in waveform templates, while the other considers only the (2,2) mode in the waveform templates. We examine sources with luminosity distance distributions ranging from 1000 to 15000 Mpc (redshift z = 0.2 - 2), studying how higher harmonics alleviate the degeneracy between luminosity distance and orbit inclination and their impact on the accuracy of source localisation. We also discuss the constraints on the number of host galaxies in both cases. Finally, on the basis of the flat $\Lambda$CDM model, we compute the Hubble constant. Thus, analysing these BBH sources enables us to better understand the influence of higher harmonics on cosmological inferences.

The structure of this paper is as follows: In section \ref{sec:2}, we introduce the higher model waveform template we used and detail the preparatory work. In section \ref{sec:3}, we show the results of parameter estimation and discuss the impact of considering higher harmonics in the waveform template on the number of host galaxies and the precision of Hubble constant calculations. In Section \ref{sec:4}, we discuss and summarize the results of this study.


\section{Waveform and methodology}\label{sec:2}

We utilize the IMRPhenomXPHM model \cite{Pratten_2021} to generate the waveforms. This waveform template is accessible within LALSuite \cite{2020asclsoft12021L}. IMRPhenomXPHM represents a frequency-domain waveform model that incorporates higher harmonics and considers the impact of orbital precession on spin-misaligned compact binary coalescences (CBCs). IMRPhenomXPHM employs a ``twisting up" technique \cite{Hannam_2014,Schmidt_2011} to construct waveforms. This approach involves the creation of two non-precession waveforms, IMRPhenomXAS and IMRPhenomXHM, within the non-inertial L-frame, which are subsequently transformed into the inertial J-frame using the precession dynamic model. In the waveform creation process of IMRPhenomXPHM, the initial step involves ``twisting up" the gravitational wave signal and decomposing it into spin-weighted spherical harmonics within the J-frame system.

\begin{equation}
h^{J} = h^{J}_{+} - ih^{J}_{\times} = \sum_{l\geq 2}\sum_{m=-l}^{l} h^{J}_{lm}{}_{-2}Y_{lm}(\theta,\phi)
\end{equation}
where,
\begin{equation}
_{-2}Y_{lm}\left(\theta,\phi\right)=\mathcal{Y}_{lm}\left(\theta\right)e^{im\phi}
\end{equation}
are the spin-weighted spherical harmonics of spin-weight -2 \cite{Goldberg1967SpinsSH},defined as in \cite{2007JCoPh.226.2359W}.

As we mentioned, the GW signal $\mu(\theta)$ is generated by IMRPhenomXPHM, where $\theta$ comprises 15 parameters, represented as $\theta = \{\mathcal{M}{c}, q, d_{L}, \theta_{JN}, \text{ra}, \text{dec}, \psi, a_{1}, a_{2}, \theta_{1}, \theta_{2}, \phi_{12},\newline \phi_{JL}, t_{c}, \phi_{c}\}$. Here, $\mathcal{M}{c}$ and $q$ denote the chirp mass and mass ratio of the two black holes, respectively. $d_{L}$ denotes the luminosity distance, while ``ra" and ``dec" describe the sky areas of the event, and $\psi$ represents the polarisation angle. $\theta_{JN}$ stands for the inclination angle at the reference frequency (set to 10 Hz in this work) for a processing system. In addition, $a_{1}$ and $a_{2}$ denote the dimensionless spin magnitudes of the two black holes. The four angles $\{\theta_{1}$, $\theta_{2}$, $\phi_{JL}$, and $\phi_{12}\}$ quantify the spin misalignment of the binary. Moreover, $t_{c}$ represents the merging time, and $\phi_{c}$ signifies the coalescence phase. In this study, we marginalise the parameters $t_{c}$ and $\phi_{c}$.

\begin{figure}[htbp]
\centering
 \includegraphics[width=0.7\textwidth]{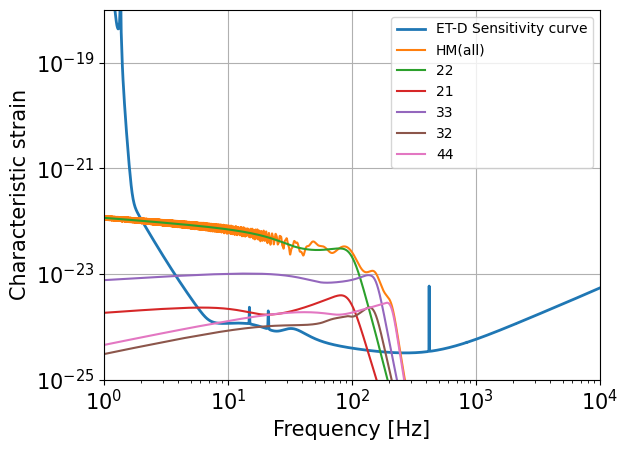}
 \caption{Comparison of the ET-D sensitivity curve with the single harmonic components of the GWs. The parameters are $M_c = 30M_{\odot}$, $q = 0.09$, $D_L = 9954$ Mpc, and $\theta_{JN} = 0.5$. The solid blue line represents the detector's characteristic strain, the solid yellow line represents the total characteristic strain of all harmonics, and the remaining solid coloured lines represent individual harmonic components.}
 \label{fig:1}
\end{figure}

As a demonstration, the signals with higher harmonics are shown in Figure \ref{fig:1}, the source parameters are $M_{c} = 30M_{\odot}$, $q = 0.09$, $D_{L} = 9954$ Mpc, $\theta_{JN} = 0.5$. The characteristic strain of single harmonics and the total characteristic strain of all harmonics are calculated. By comparing with the sensitivity curve of ET, we find that the $(3, 3)$ mode is significant, and the other modes also matter. Therefore, for the next generation of ground-based gravitational wave detectors, ET, the inclusion of higher harmonics can provide additional information from observational data.

The signals are injected into the Gaussian noise generated with the design sensitivities of ET \cite{Maggiore_2020ET} and CE \cite{2019BAAS...51g..35R}.  To generate posterior samples, we employed a fast sampler, Dynesty \cite{Speagle_2020}. In parameter estimation, we fix the spin parameters $\{a_{1}$, $a_{2}$, $\theta_{1}$, $\theta_{2}$, and $\phi_{12}\}$ to the injected values. Although this approach may result in a narrower conditional posterior to a fully marginalised posterior, it adequately serves the purpose of investigating distance-inclination degeneracy.

The Bayesian parameter estimation package Bilby \cite{Ashton_2019} is used to estimate the source parameters and compare the localisation inference capabilities of the different modes. The standard Gaussian noise likelihood function $\mathcal{L}$ is used to analyse the strain data $d_{k}$ based on the following source parameters: $\theta$ \cite{Romero_Shaw_2020}

\begin{equation}
    \ln{\mathcal{L}\left(d | \theta\right)} = -\dfrac{1}{2}\sum_{k}\{\dfrac{[d_{k}-\mu_{k}\left(\theta\right)]^{2}}{\delta^{2}_{k}} + \ln{\left(2\pi\delta^{2}_{k}\right)}\}
\end{equation}

For the injected signals, the prior distribution for the luminosity distance is assumed to be uniform within the range of 1000-30000 Mpc. The chirp mass is 30, the mass ratio is random selected from 0.05 to 0.1. We then employ IMRPhenomXPHM with higher harmonics and only with (2,2) mode to do the parameter estimation. This allows us to evaluate whether higher harmonics can mitigate the degeneracy between luminosity distance and inclination angle. To quantify the difference between the two probability distributions of priors and the estimated results, we employ the Hellinger distance. The Hellinger distance between two distinct probability distributions $P$ and $Q$ is defined as: $\text{Hellinger}(P, Q) = \sqrt{1 - \sqrt{\sum_{i} \sqrt{p_i \cdot q_i}}}$ \cite{Hellinger1909,Leandro_2022}. This metric measures the similarity between the probability distributions.

In addition, the luminosity distance of the source at redshift $d_{L}$(z) is given by the following formula:

\begin{equation}
d_{L}\left(z\right)=c\left(1+z\right)
\int_0^z \dfrac{dz^{'}}{H(z^{'})}\
\end{equation}

Where $c$ is the speed of light and $H(z)$ is the Hubble parameter describing the expansion rate of the universe at redshift z. In this study, we utilize the flat $\Lambda$CDM model as the base model, and $H(z)$ has the following form:

\begin{equation}
H\left(z\right)=H_{0}\sqrt{\Omega_{m}\left(1+z\right)^3+1-\Omega_{m}}
\end{equation}

Where the equation of state (EoS) parameter $w$ for dark energy is assumed to be - 1, and $\Omega_{m}$ is the current matter density parameter \cite{Frieman_2008}. The base values of the remaining parameters were selected as Planck 2018 \cite{2020} $TT,TE,EE+lowE$ results, where $\Omega_{m}=0.3166$, $H_{0}=67.27$ km/s/Mpc.

\section{Parameter estimation results}\label{sec:3}

In this section, we first show the parameter estimation results with or without higher modes. Then, combining the localisation of the sources, we constrain the host galaxies of GW events, and infer the Hubble constant by the luminosity distances of GW events and redshifts of host galaxies.

\subsection{Comparison of luminosity distance posterior distribution}

\begin{figure}[htbp]
\centering
\includegraphics[width=0.8\textwidth]{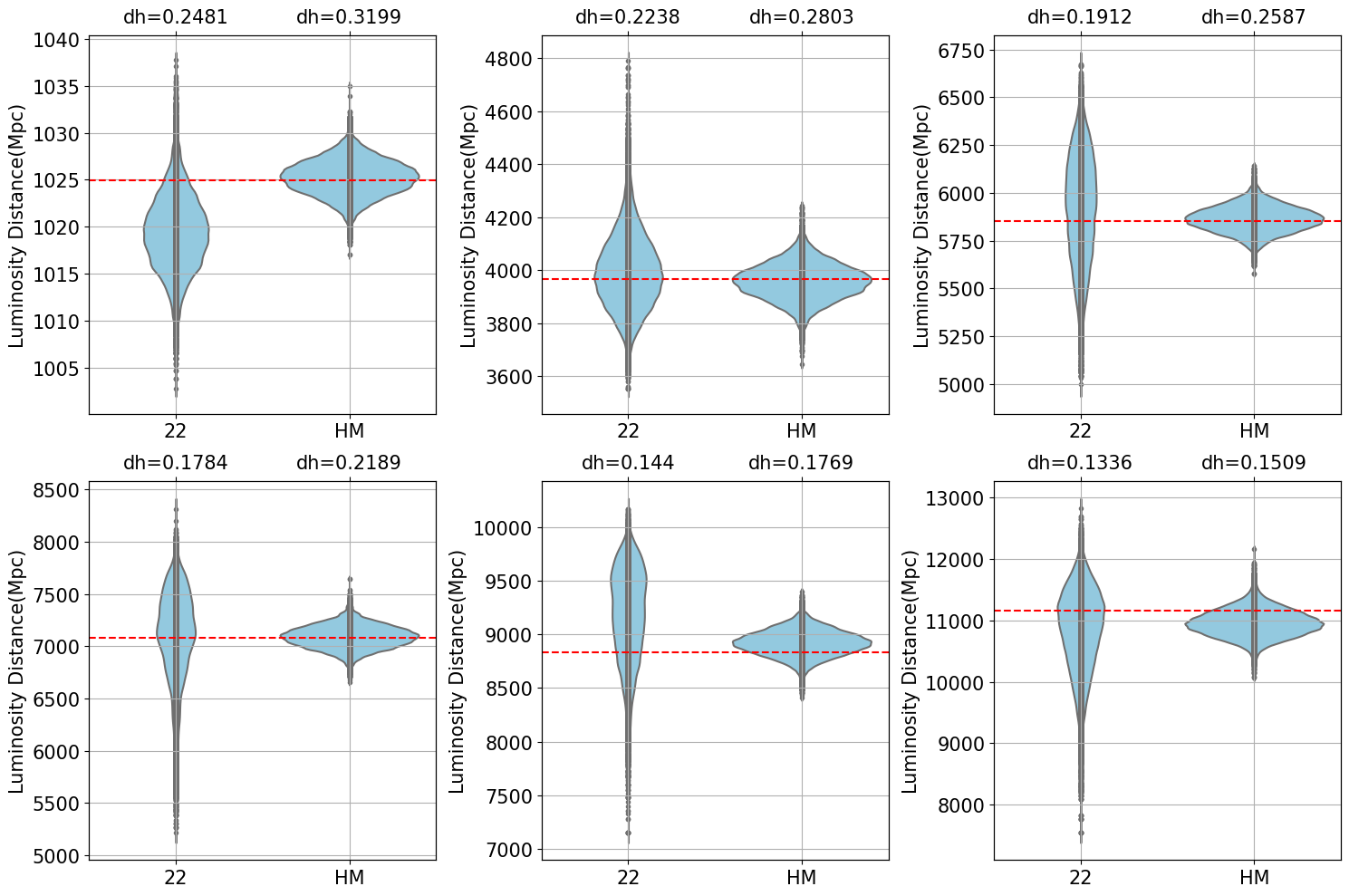}
\caption{The posterior distributions of the luminosity distance for different injected distances. The red line in each subfigure represents the injected distance for each case. The box in the middle of the violin plots represents the upper and lower quartiles of the posterior distributions. For the three subplots in the top panel, the mass ratios are 0.16, 0.13, and 0.09, while for the three subplots in the bottom panel, the ratios are 0.11, 0.17, and 0.09.}
\label{fig:2}
\end{figure}

\begin{figure}[htbp]
\centering
 \includegraphics[width=0.75\textwidth]{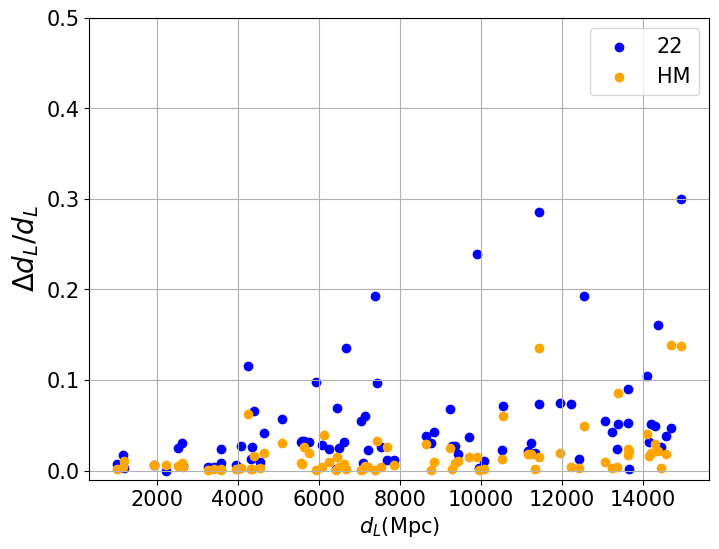}
 \caption{Confidence interval of the luminosity distance divided by the luminosity distance $(\Delta d_{l}/d_{l})$ for 80 random sources at different luminosity distances for both the HM cases (orange dots) and (2,2) modes (blue dots). The other parameters are set randomly.}
 \label{fig:3}
\end{figure}

Figure  \ref{fig:2} shows the posterior distribution of luminosity distances between (2,2) mode and higher harmonics at different distances. The violin plots in Figure \ref{fig:2} show the posterior distribution of luminosity distances at different distances. With an increase in the luminosity distance, the interval of the $d_{L}$ under the two methods gradually increases. However, when the higher harmonics are considered, the 90\% confidence interval of the BBH merger event distance is narrower than that when only the (2,2) mode is considered, resulting in an improvement of around 60\%. For distant sources, the inclusion of higher harmonics significantly improves the accuracy of the luminosity distance estimation, and the posterior distribution confidence interval of the luminosity distance slightly increases. Conversely, when higher harmonics are not considered, the confidence interval range also increases significantly with luminosity distance.

In addition, we calculate the Hellinger distance to quantify the difference between the prior distribution and the posterior distribution. The upper part of Figure \ref{fig:2} shows the Hellinger distance values at different luminosity distances.

Figure \ref{fig:3} shows the parameter estimation result of the luminosity distance estimation for BBH events under (2,2) mode and higher harmonics by injecting 80 randomly-assigned parameter sources. These injected BBHs have luminosity distance settings ranging from $1000$ to $15000$ Mpc. For BBH events considering higher harmonics, $\Delta d_{L}$ remains below 10\%, while for events considering only (2,2) mode, $\Delta d_{L}$ exceeds 10\%. At closer luminosity distances, the estimation results of both methods are accurate; however, as the luminosity distance increases, the differences between the two methods gradually become clear. Considering higher harmonics results in a more pronounced improvement in the uncertainty of the luminosity distance parameter estimation for BBH events by 2-15 times.  In contrast, for BBH events considering only the (2,2) mode, their $\Delta d_{L}$ generally exceeds that of the former case but also shows a diminishing improvement with increasing distance. This is consistent with the decrease in the signal-to-noise ratio with increasing distance.

\subsection{Degenerate remission}

\begin{figure}[htbp]
\centering
 \includegraphics[width=0.7\textwidth]{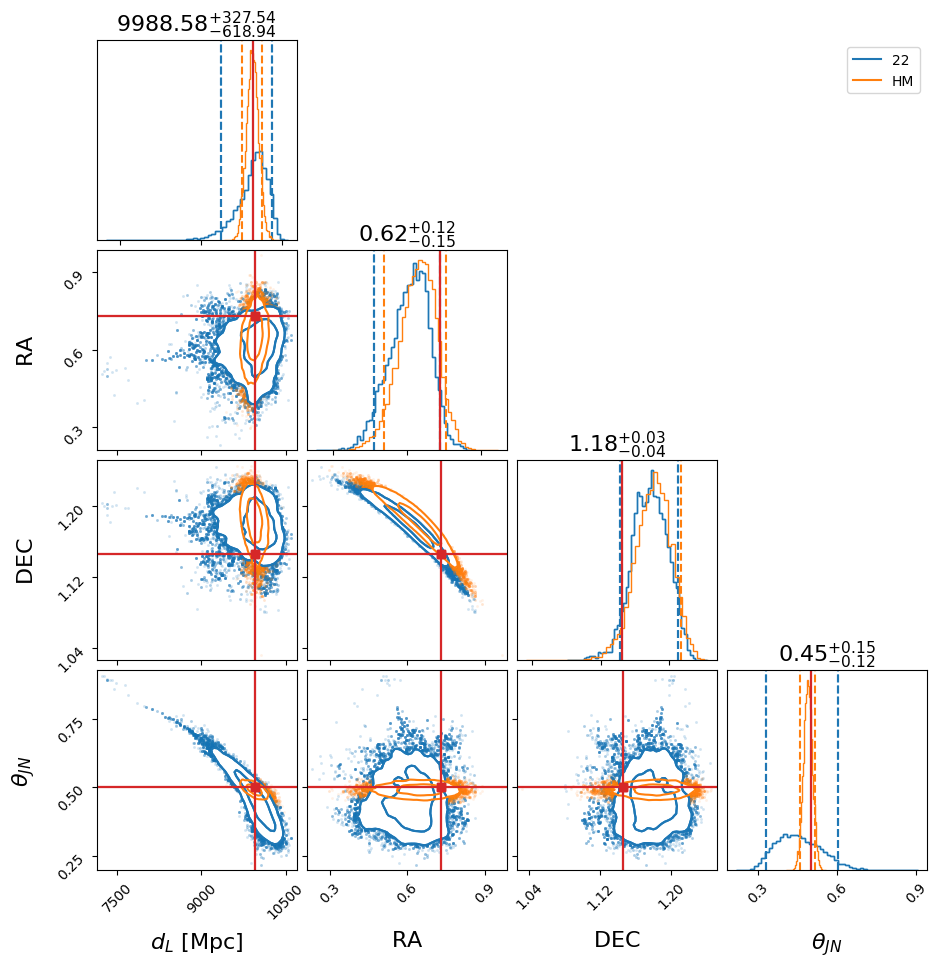}
 \caption{Comparison of the parameter estimation result of BBH using higher harmonics and only (2,2) mode. The injected luminosity distance is 9954 Mpc.}
 \label{fig:4}
\end{figure}

Figure \ref{fig:4} presents the parameter estimation results for gravitational wave sources at a luminosity distance of 9954 Mpc. When considering only the (2,2) mode, a noticeable degeneracy between the orbital inclination and the luminosity distance is observed, which decreases the accuracy of the luminosity distance parameter estimation. However, the inclusion of higher harmonics effectively reduces this degeneracy, thereby improving the accuracy of the luminosity distance parameter estimation. In addition, in our parameter estimation results, we observed a significant improvement in localisation precision compared to existing gravitational wave detectors for both methods, with calculated sky areas ranging within several tens of square degrees. However, concerning the parameter estimation results for CE and ET, the inclusion or exclusion of higher harmonics in the templates does not significantly improve the precision of source localisation.

These results emphasise the importance of introducing higher harmonics into waveform templates when studying gravitational wave sources at different distances. Furthermore, the research results demonstrate the importance of introducing higher harmonics in future third-generation detectors such as CE and ET. By employing this method, it is possible to more accurately constrain the number of host galaxies, thereby utilizing gravitational waves as cosmological probes to measure the Hubble constant $H_{0}$.

\subsection{Constraints on the number of host galaxies}

\begin{figure}[htbp]
\centering
 \includegraphics[width=0.72\textwidth]{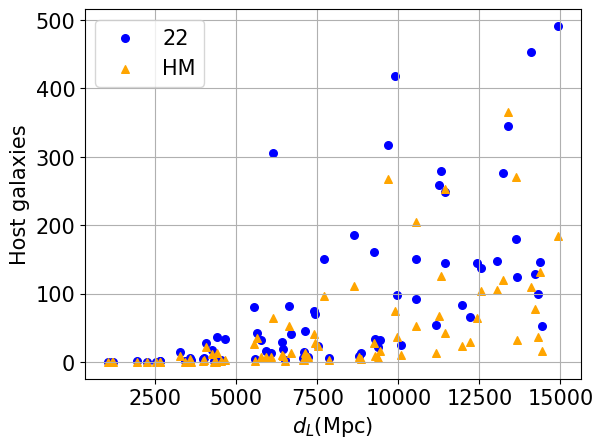}
 \caption{Comparing across different modes, we examine how the 80 sources constrain the host galaxies at various luminosity distances. The HM mode is denoted by orange, while the (2,2) mode is represented by blue. All other parameters are randomly assigned.}
 \label{fig:5}
\end{figure}

We assume that galaxies are evenly distributed throughout the comoving volume \cite{2020arXiv201014732W}, allowing us to estimate the number of galaxies within a given volume. Using the inferred sky areas and distance inferences, we calculate the possible comoving volume of potential host galaxies. The number density of galaxies in the universe is $0.02 Mpc^{-3}$ \cite{Barausse_2012,2020arXiv201014732W}. Multiplying this density by volume provides an estimate of the number of potential host galaxies. As shown in Figure \ref{fig:5}, the inclusion of higher harmonics in a binary black hole system leads to a gradual decrease in the number of potential host galaxies as the luminosity distance increases compared with a (2,2) mode binary black hole system.


\subsection{Calculation of the Hubble constant}

\begin{figure}[htbp]
\centering
 \includegraphics[width=0.72\textwidth]{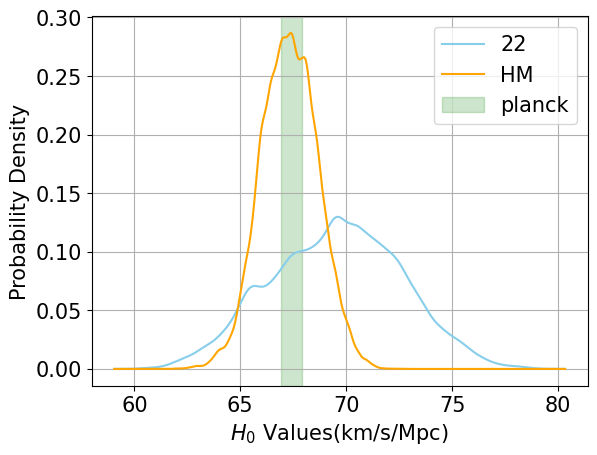}
 \caption{Comparison between the estimation of $H_{0}$ for dark sirens originating from HM BBHs and those from (2,2) mode BBHs.}
 \label{fig:6}
\end{figure}

Finally, we perform a simple evaluation to investigate the effect of higher harmonics on the accuracy of the $H_{0}$ inference results. Using the results of parameter estimation, we can determine the target sky areas where the light source is located. We then select host galaxies from the SDSS DR18 \cite{Almeida_2023} catalog, and weight the redshifts of these host galaxies to obtain the redshifts of the source. We can then calculate the Hubble constant range using the standard cosmological model and the 90\% confidence interval for the redshift and luminosity distances of the host galaxy. Our calculation results are displayed in Figure \ref{fig:6}. The orange curve represents the $H_0$ inferred when higher harmonics are considered in the template, whereas the blue curve represents the $H_0$ inferred when only the (2,2) mode is considered in the template. It is evident from the figure that the higher harmonics BBH source can provide a more tightly constrained $H_{0}$ estimate as a dark sirens. The distribution in Figure \ref{fig:6} is normalised such that its area is equal to 1.

Figure \ref{fig:6} illustrates the calculation results. In the case where higher harmonics are considered in the template, the result for $H_{0hm}$ is $67.31^{+2.19}{-2.13}$ km/s/Mpc, whereas when only the (2,2) mode is considered, the result for $H_{022}$ is $69.61^{+4.89}_{-5.16}$ km/s/Mpc. The green shadow corresponds to the 68\% confidence interval calculated from the Planck 18 data. Notably, when considering higher harmonics, our constraint on $H_{0}$ becomes more stringent and precise. The ratio of $\Delta H_{0}/H_{0}$ of 3\% for the (2,2) mode and a ratio of $\Delta H_{0}/H_{0}$ of 0.6\% for higher harmonics.

\section{Discussion and conclusion}\label{sec:4}

In cosmology, the Hubble constant can be measured by different methods; however, there are significant discrepancies in the results obtained by different methods, known as the Hubble tension \cite{freedman2017cosmology}. To address this issue, independent measurements of the Hubble constant are required. Using BBHs as standard sirens to measure the $H_{0}$ is a highly effective method, with the error in the Hubble constant depending on the errors in the luminosity distance and redshift; where the luminosity distance is degenerate with some parameters like as the inclination angle, and the redshift is constrained by the localisation of the host galaxy.

In this study, we investigated whether considering higher harmonics or just the (2,2) mode in templates could improve the luminosity distance and localisation measurements of BBH events. Our results indicate that incorporating higher harmonics effectively alleviates the degeneracy between the luminosity distance and inclination angle, thereby enhancing the precision of $d_{L}$ in parameter estimation. In addition, we found that including higher harmonics in the templates does not significantly improve the source localisation precision. However, compared with the results from LIGO/Virgo, there is a substantial improvement in the localisation accuracy. Furthermore, we computed the number of host galaxies and observed a significant decrease in their number at farther distances, implying reduced redshift errors and improved accuracy in Hubble constant calculations. Lastly, we calculated the Hubble constant and found that considering higher harmonics in the templates improves its precision by an order of magnitude.

In addition, to fully exploit the potential of standard sirens for inferring cosmological parameters, a more complete catalogue of galaxies is required. In the future, with the introduction of the next generation of larger field telescopes such as the China Space Station Telescope (CSST) \cite{csst}, we will be able to obtain a more complete catalogue of stars. This will improve the accuracy of the cosmological parameter estimation results. It is also worth noting that we set the cutoff frequency of the compact binary merger to 10 Hz (which is consistent with the lowest aLIGO cutoff frequency \cite{Martynov_2016} ), which means that contributions from all modes below 10 Hz are discarded. In the next generation of observation equipment, the cutoff frequency of CE will be extended to 5 Hz \cite{Hall_2021CE} and that of ET will be extended to 1 Hz \cite{2012CQGra..29l4013S}, which means that expanding the observation band to a lower frequency may yield better results.

The results presented in this paper highlight the potential of treating gravitational wave sources as ``dark sirens" to enhance our understanding of the expansion history of the universe. While our research focuses on third-generation ground-based gravitational wave detectors, our findings can also be extended to space-based detectors \cite{Marsat_2021,gong2023including}. By using the ``dark sirens" with higher harmonics, we may measure the Hubble constant more accurately, and thus gain a deeper understanding of the history and evolution of the universe.

\acknowledgments

This work was supported by the National Key R\&D Program of China (Grant Nos. 2021YFC2203002), and the National Natural Science Foundation of China (Grant No. 12173071). We thank Deng-Hui Nie for his help in this study. He generously shared his expertise and experience, providing valuable advice on determining host galaxy redshifts. This work made use of the High-Performance Computing Resource in the Core Facility for Advanced Research Computing at Shanghai Astronomical Observatory.



\bibliographystyle{JHEP}

\providecommand{\href}[2]{#2}\begingroup\raggedright\endgroup



\end{document}